\documentclass[12pt]{article}
\usepackage{jheppub}
\usepackage{amsmath,amssymb}

\makeatletter

\@addtoreset{equation}{section}
\makeatother

\newcommand{\beq}{\begin{equation}}
\newcommand{\eeq}{\end{equation}}

\begin{document}

\baselineskip=18pt  
\baselineskip 0.7cm

\begin{titlepage}

\setcounter{page}{0}

\renewcommand{\thefootnote}{\fnsymbol{footnote}}

\begin{flushright}
\end{flushright}

\vskip 1.5cm

\begin{center}
{\LARGE \bf
Large $N$ Duality, Mirror Symmetry,

 \vskip 0.5cm
and a
\vskip 0.5 cm
Q-deformed A-polynomial for Knots}

\vskip 1.5cm 

{\large
Mina Aganagic$^{1,2}$ and Cumrun Vafa$^3$ 
\\
\medskip
}

\vskip 0.5cm

{
\it
$^1$Center for Theoretical Physics, University of California, Berkeley, CA 94720, USA\\
\medskip
$^2$Department of Mathematics, University of California, Berkeley, CA 94720, USA\\
\medskip
$^3$Jefferson Physical Laboratory, Harvard University, Cambridge, MA 02138, USA\\
\medskip
}

\end{center}

\centerline{{\bf Abstract}}
\medskip
\noindent
We reconsider topological string realization of $SU(N)$ Chern-Simons theory on $S^3$. At large $N$, for every knot $K$ in $S^3$, we obtain a polynomial $A_K(x,p;Q)$ in two variables $x,p$ depending on the t'Hooft coupling parameter $Q=e^{Ng_s}.$  Its vanishing locus is the quantum corrected moduli space of a
special Lagrangian brane $L_K$, associated to $K$, probing the large $N$ dual geometry, the resolved conifold.
Using a generalized SYZ conjecture this leads to the statement that for every such Lagrangian brane $L_K$ we get a distinct mirror of the resolved conifold given by $uv=A_K(x,p;Q)$.
Perturbative corrections of the refined B-model for the open string sector on the mirror
geometry capture BPS degeneracies and thus the knot
homology invariants. Thus, in terms of its ability to distinguish knots, the classical function $A_K(x,p;Q)$ contains at least as much information as knot homologies. 
In the special case when $N=2$, our observations lead to a physical explanation of the generalized (quantum) volume conjecture.  
Moreover, the specialization to $Q=1$ of $A_K$ contains the classical A-polynomial
of the knot as a factor. 

\end{titlepage}
\setcounter{page}{1} 

\section{Introduction}
The study of knot invariants has a long history in mathematics.  With the introduction of Jones polynomials \cite{RJ}
and their physical interpretation  in terms of Chern-Simons theories \cite{Witten:1988hf}
and their embedding in topological strings \cite{Witten:1992fb} this has led to an interesting area of interaction between mathematics
and physics in modern times. With the advent of string dualities new perspective was gained
on this relation.  Large $N$ dualities for topological strings \cite{GV} lead
to new predictions for HOMFLY polynomials.  In particular the interpretation of topological
strings as computing BPS degeneracies \cite{OV} has led to integrality predictions for colored
HOMFLY polynomials.  Furthermore, consideration of symmetries of the problem, leads
to an interpretation of the Khovanov invariants \cite{K} and knot homologies \cite{KR1, KR2}, in terms of further decomposition of this BPS Hilbert space with respect to an additional $U(1)$ symmetry \cite{Gukov:2004hz} (see also the recent work \cite{wittenk}).
This refinement in turn relates to Nekrasov deformation of topological strings \cite{Gukov:2007tf}. The deformation of the topological string also leads to a refinement of Chern-Simons theory recently discovered in \cite{MS}.  

In order to compute knot invariants using topological strings, one needs to get a good
handle on the Lagrangian brane $L_K$ associated to a knot, after the large $N$
transition.   Before the large $N$ transition, the geometry of the Calabi-Yau is
$T^*S^3$ and $L_K$ is obtained from the knot $K$ by the conormal bundle
construction \cite{OV}.  However, for general knots it
has been difficult to use the large $N$ duality to make concrete predictions about
the knot invariants, as one needs a simple description of the $L_K$ after
the large $N$ transition to the resolved conifold.  For unknot this can be done explicitly, because it turns out
that there is a mirror symmetry \cite{HV} which is compatible with the Lagrangian
associated with the unknot
and which maps the computation of the knot invariants to easy computations on the mirror.
Until now it has been difficult to apply this idea to general knots
and use topological strings as a practical method for computations of all knot invariants.
One aim of this paper is to remedy this gap.  We formulate a remarkable
generalization of mirror symmetry in the spirit of SYZ \cite{SYZ}, which allows us
to compute, for the  resolved conifold, a distinct mirror for each choice of knot $K$.  We find that for every
knot $K$ in $S^3$, the leading large $N$ limit of the knot invariants, leads to a polynomial
$$
A_K(e^x,e^p,Q)=0,
$$
which captures the moduli of $L_K$ corrected by disc instantons.
The polynomial $A_K(e^x,e^p,Q)$ captures the mirror geometry in the spirit of SYZ. It determines 
the mirror Calabi-Yau  $Y_K$, as a hypersurface
$$
A_K(e^x,e^p,Q)=uv.
$$

The polynomial $A_K(e^x,e^p,Q)$ can be viewed as an invariant of the knot. Mirror symmetry and large N-duality relate 
the computation of $SL(N)$ knot invariants to computation of open topological string amplitudes on the mirror Calabi-Yau $Y_K$, with a brane mirror to $L_K$. 
Nekrasov deformation of the topological string on $Y_K$ is relevant for categorified knot invariants \cite{Gukov:2004hz}. 
But, topological string amplitudes carry no more information about distinguishing Calabi-Yau than the classical geometry. 
This leads us to conclude that categorification carries no additional information for distinguishing knots beyond that captured in the polynomial $A_K(e^x,e^p,Q)$.

In a seemingly unrelated development, another polynomial for knots was introduced to physics in 
\cite{Gukov:2003na}. This is the A-polynomial of the knot, which characterizes flat
$SL(2)$ holonomies on the knot complement (mathematical work on the A-polynomial originated in \cite{Apol}).
In \cite{garo,Gukov:2003na,GT} 
the generating polynomial for colored Jones
was studied and conjectured to satisfy a difference equation 
(see also \cite{Garoufalidis:2011dx,Gukov:2011qp}).   These difference
equations can be viewed as a quantum operator associated to the  A-polynomial. This structure
is reminiscent of difference equations satisfied by open topological string amplitudes
\cite{IH} and in fact in some cases it was checked that the corresponding objects are canonically
isomorphic, namely that a quantum deformation of the A-polynomial maps to a local
geometry which in turn captures the topological string amplitudes \cite{2009ForPh..57..825D,Dijkgraaf:2010ur}.  

The secondary aim of this paper is to explain these observations in the context of the broader
picture of the relation between knots and topological strings and
to extend these observations to higher rank $N$.  We find that the $A_K(e^x,e^p,Q)=0$ geometry
naturally computes the generating function relevant for the quantum volume conjecture.
Moreover,
the polynomial that captures
the mirror associated to each knot turns out to be a $Q$-deformed version of the
classical A-polynomial of the knot, where $Q$ is related to the large $N$ `t Hooft parameter.
More precisely as $Q\rightarrow 1$ the mirror geometry contains the A-polynomial
of the knot as a factor.
 In this context, the fact that
$A_K$ is related to an operator annihilating the partition function of the knot, follows
from general properties of open topological strings \cite{IH}.  

The plan for this paper is as follows:  In section 2 we recall
some basic aspects of mirror symmetry picture according to SYZ \cite{SYZ} and introduce
an extension of it which is relevant for local non-compact Calabi-Yau. In section 3 we review the relation of Chern-Simons theory to topological strings on $T^*S^3$ and the Lagrangian $L_K$ associated
to a knot $K$.  We also recall aspects of  large $N$ duality.  In section 4, using large $N$ duality, we sum up disk instanton corrections to the moduli space of the brane on the resolved conifold, and discover a family of mirrors, one for each knot $K$.
In section 5 we review the A-polynomial for knots and relations to volume and AJ conjectures and derive these conjecture in the setup of topological string.  Furthermore
we interpret our Q-deformed A-polynomial as a generalization of
the volume conjecture.
In section 6 we give examples. 
In  section 7, we discuss the relation between our polynomial and
a similar one which has been proposed in the context of knot contact homology.
 In section 8 we make
 some concluding remarks
and suggestions for future direction of research.

\section{Mirror Symmetry in the Local Case}
In this section we review certain features of mirror symmetry for non-compact Calabi-Yau manifolds. We will focus on the local toric Calabi-Yau three-folds, for which mirror symmetry is understood the best. This is also the case that naturally fits with knot theory, as we will see.

It is known \cite{HV,HIV} that for the case of non-compact toric Calabi-Yau the mirror Calabi-Yau 3-fold takes the form
\beq\label{mirrorCY}
F(e^x,e^p)=uv
\eeq
where $(e^x,e^p,u,v)\in {\bf C}^*\times {\bf C}^* \times {\bf C}\times {\bf C}$.  In particular the main data of the Calabi-Yau
is captured by the curve 
\beq\label{mirrorcurve}
F(e^x,e^p)=0.
\eeq

As discussed in \cite{MV} this curve can be viewed as the moduli space of a canonical special
Lagrangian brane.  These special Lagrangian submanifolds have the topology of $S^1\times {\bf R}^2$ and generalize the original constructions of Harvey and Lawson, and Joyce \cite{HL, Joyce}. By a theorem of McLean \cite{McLean}, a special Lagrangian with a  $b_1=1$ has a one real dimensional moduli space coming for the deformations, and this is accompanied by the moduli of a flat bundle. If we parameterize the moduli space by 
$x=r+i\theta$, $r$ comes from geometric deformations of the Lagrangian, and $\theta=\oint_{S^1} A$ from the choice of flat connection on the $S^1$.
The moduli space receives corrections from holomorphic disks ending on the brane. The disk amplitude $W(x)$ depends holomorphicaly on $e^{-x}$. The area of the disk is determined by the position $r$ of the Lagrangian. 
Mirror symmetry relates the disk instanton corrections on the A-model side
to classical periods of the mirror. As explained in \cite{MV}, in the mirror, the disk amplitude $W(x)$ is the integral of the holomorphic threeform $\Omega$ on a chain whose boundary
is the Lagrangian manifold, which for Calabi-Yau manifolds of the form \eqref{mirrorCY} descends to a one form $pdx$ on the curve \eqref{mirrorcurve}.:
$$
W(x) = \int^x p(x) dx.
$$
The disk instanton corrected moduli space of these Lagrangian submanifolds is captured
by a choice of a point on the mirror curve $ F(e^x,e^p)=0$.  In other words we can parameterize the moduli 
space of the brane by the choice of $x$ and find $p=p(x)$, such that $F(e^x,e^p)=0$.

A simple example of this is the resolved conifold geometry given by 
$$X={\cal O}(-1)\oplus {\cal O}(-1)\rightarrow {\bf P}^1.$$
Let $Q=e^{-t}$
where $t$ denotes the complexified Kahler class of ${\bf P}^1$.  Then the mirror geometry is captured by the curve \cite{HV}
$$F(e^x,e^p)=1+e^x+e^p+Qe^{x+p}=0.$$
This curve can, in turn, be viewed as the moduli space of the special Lagragian branes on the resolved conifold.
At one point along their moduli the $S^1$ of the Lagrangian is identified with the equator of ${\bf P}^1$, while the ${\bf R}^2$ is a suitable
subspace of the ${\cal O}(-1)\oplus {\cal O}(-1)$ vector bundle over it.  Note that the classical moduli space of this
Lagrangian brane has singularities, where the $S^1$ shrinks to a point. The singularities are smoothed out in the quantum theory by the disc instantons.
In particular the geometry $F(e^x,e^p)=0$ has no singular points, and this is a reflection of the fact
that the classical geometry of the mirror manifold already captures the disk instanton sum on the
A-model side.

In the SYZ formulation for mirror symmetry \cite{SYZ}, in the compact case, one considers on the A-side a special
Lagrangian with the topology of $T^3$.  This has a 3-complex dimensional moduli space.  The mirror
is identified as the quantum corrected moduli space, where the quantum corrections include the disc
instantons.  In the non-compact case it is natural to consider extensions of this where the special
Lagrangian has a different topology.  In particular from the discussion we just had, it is clear that a natural
topology for the special Lagrangian to consider on the A-side is that of ${\bf R^2}\times S^1$ and in this context
the curve $F(e^x,e^p)=0$ is naturally identified as the disk instanton corrected moduli space of the Lagrangian brane.\footnote{
The Lagrangian branes with the topology of ${\bf R}^1\times T^2 $ have also been considered in this context in \cite{Conan1,Conan2} for toric Calabi-Yau manifolds and 
relating the SYZ mirror picture with the mirror picture of \cite{HV}. } 

However the SYZ construction of mirror geometry admits a more general interpretation.  Namely,
we can consider {\it an arbitrary} special Lagrangian manifold $L$ with the topology of ${\bf R}^2\times S^1$.  For each
such choice, by summing up disk instanton corrections, we obtain a mirror geometry. The disc corrected moduli space is
given by the curve 
$$F_L(e^x,e^p)=0,
$$
which now depends on the Lagrangian $L$.  Thus we end up with {\it infinitely} many mirrors for
each toric manifold.  In particular ${\cal O}(-1)\oplus {\cal O}(-1)\rightarrow {\bf P}^1$ can have infinitely many mirrors in
the SYZ sense.  The SYZ conjecture in this case becomes the statement that no matter which one of these
choices of the B-model geometry we pick, we should nevertheless get identical topological string amplitudes.  Moreover the open string amplitudes involving
the Lagrangian $L$ on the A-model side can be captured as the canonical
brane on the B-model side in this mirror geometry -- this brane is the brane that corresponds to a choice of a point on the curve, and the subspace $v=0$ of the mirror Calabi-Yau
$$F_L(e^x,e^p)=uv.
$$

In the next section, we will explain how, using the relation of topological string on $X={\cal O}(-1)\oplus {\cal  O}(-1)\rightarrow {\bf P}^1$ to Chern-Simons theory on $S^3$, and large $N$ dualities, we can derive, for each 
knot $K$ in the $S^3$ a distinct mirror $Y_K$ mirror to $X$ depending on the knot. The topological string on each such $Y_K$ should be the same. 
\section{Topological String and Chern-Simons Theory}

Consider topological A-model on 
$$
X^v= T^* S^3
$$
with
$N$ Lagrangian D-branes on the $S^3$. The partition function of the open A-model topological string is the $G=SU(N)$ Chern-Simons partition function
on the $S^3$ \cite{Witten:1992fb},
$$
Z^{top}(X, g_s) = Z_{CS}(S^3, k)
$$
where the topological string coupling $g_s$ coincides with the Chern-Simons coupling constant,
$$g_s = \frac{2 \pi i}{k+N}.$$
A nice way to introduce knot invariants in this setup is to associate a particular
Lagrangian $L_K$ to the knot \cite{OV}, so that
$$L_K \cap S^3 = K.
$$
This is the conormal bundle construction, which basically means choosing the ${\bf R}^2$ planes of $L_K$
to be in the cotangent fiber direction, and be orthogonal (relative to symplectic pairing)
to the tangent to the knot. 
Adding a D-brane on $L_K$ in the topological string corresponds to inserting
$$
Z^{top}(X, L_K, x) = \langle det(1\otimes 1 - e^{-x} \otimes U)^{(-1)}\rangle_{S^3}  
$$
in Chern-Simons theory, where $U$ is the holonomy along the knot, $U = P e^{i \oint_K {\cal A}}$. 
Since the Lagrangian $L_K$ has topology ${\bf R}^2 \times S^1$, the theory on the brane has one modulus, which we will denote by $x$. The determinant captures the effect of integrating out the bifundamental scalar field of mass $x$, corresponding to the string stretching between the $S^3$ and $L_K$ \cite{OV}.  We can rewrite above as\footnote{This follows from the Cauchy formula in the theory of symmetric functions \cite{macdonald}.}
\beq\label{TopS}
Z^{top}(X, L_K, x)=  \sum_{n=0}^{\infty} \langle Tr_{R_n} U \rangle_{S^3} \; e^{-n x } 
\eeq
On the right hand side, the sum runs over $R_n$, the totally symmetric representations of $SU(N)$ with $n$ boxes. Let 
$$
\langle  Tr_{R_n} U\rangle_{S^3} = H_{n}(K),
$$
denote the expectation value of the Wilson loop along the knot $K$, colored by representation $R_n$ in the $SU(N)$ Chern-Simons theory.

For general $G=SU(N)$, the topological string partition function defines a wave function
\beq\label{TW}
Z^{top}(X,L_K, x)= \sum_n H_n(K) \; e^{-nx}=\Psi_{K}(x)
\eeq
which is the exact partition function of the D-brane on $L_K$, 
$$
\Psi_{K}(x)=\Psi_{K}(x,g_s,N)
$$
depending on $g_s$ and $N$. It is a wave function because $L_K$ is a non-compact three manifold, with a $T^2$ boundary; $x$ here corresponds to the holonomy around the $S^1$ in $T^2$ that remains finite in the interior of $L_K={\bf R}^2\times S^1$.

\subsection{Large $N$ duality}

It is conjectured in \cite{GV} that, if we have $N$ branes wrapped on $S^3$ in $X^v=T^*S^3$,
at large $N$, the geometry undergoes a transition
where $S^3$ shrinks and the $N$ Lagrangian branes disappear, leaving
behind a blown up ${\bf P}^1$.  In other words it is conjectured
that the $SU(N)$ Chern-Simons theory is equivalent at large $N$ to the 
topological string on 
$$X={\cal O}(-1) \oplus {\cal O}(-1)\rightarrow {\bf P}^1.$$
where the t'Hooft coupling of Chern-Simons theory becomes the size of the ${\bf P}^1$ in $X$,  
$$
Q=exp(-t),\qquad t=Ng_s.
$$ 
The geometry of both $X$ and $X^v$ are cones over $S^2\times S^3$.
The geometric transition shrinks the $S^3$ at the apex of the cone in $T^*S^3$ and replaces it by the ${\bf P}^1$, 
leaving the geometry far from the tip untouched.

Now consider the noncompact branes $L_K$ we constructed in the previous section, one for each knot $K$ in the $S^3$.
This Lagrangian gets pushed through the transition to a Lagrangian $L_K$ on $X={\cal O}(-1) \oplus {\cal O}(-1)\rightarrow {\bf P}^1$. We will abuse notation, and denote the Lagrangians on both $X$ and $X^v$ by the same letter, $L_K$. The topology of the Lagrangian is unchanged by the transition -- $L_K$ is still ${\bf R}^2\times S^1$. To see this, note that we could have used the real modulus on the Lagrangian to push $L_K$  away from the $S^3$  before the transition -- this way, the Lagranian can pass through the transition while remaining far from the apex of the cone where the geometric transition takes place. Thus, large $N$ duality and the geometric transition gives us a way to construct a family of Lagrangians $L_K$ of the topology of ${\bf R}^2\times S^1$ in ${\cal O}(-1) \oplus {\cal O}(-1)\rightarrow {\bf P}^1$, one for each knot $K$ in the $S^3$. 

The large $N$ duality also gives a way to compute topological string amplitudes on $X$ with and without branes. The partition function of Chern-Simons theory on the $S^3$, as shown in \cite{GV}, is the same as the closed topological string partition function of $X$. Similarly, the partition function of the theory on $T^*S^3$ with a D-branes on $L_K$ gets related by duality to topological string on $X$ with an A-brane on $L_K$. After the transition, one is computing open topological string amplitudes with contributions from holomorphic maps with boundaries on $L_K$. The $N$ of Chern-Simons theory is captured by $Q = e^{-t} = q^N$. $Q$ keeps track of the degrees of the holomorphic maps -- their winding number around the ${\bf P}^1$. In particular, we can use Chern-Simons theory to sum up the disk instanton corrections to the moduli space of the Lagrangian brane on $L_K$ and, in this way derive the geometry of the mirror, as will
be discussed in the next section.

\section{Mirror Symmetry and Knot Invariants}

We will now combine the various ideas discussed in the previous sections.
We start with the $T^*S^3$ geometry and wrap $N$ Lagrangian branes on the $S^3$.  This leads
to $SU(N)$ Chern-Simons gauge theory on $S^3$.  We consider a knot $K$ on $S^3$ and let $L_K$ be
the associated special Lagrangian brane which intersects the $S^3$ on $K$. %
As reviewed  in section 3 at large $N$
the $T^*S^3$ undergoes the geometric transition to the resolved conifold geometry with the modulus
of the blown up ${\bf P}^1$ given by $t$ where $Q=e^t$.  Moreover the $L_K$ will
be represented by a special Lagrangian submanifold, which by abuse of notation we still denote by $L_K$.
Note that $L_K$ has the topology of ${\bf R}^2\times S^1$ and so the discussion of the previous
section applies. 
The theory has a mirror geometry which is captured by a curve which we
denote by 
\beq\label{mirror}
F_K(e^x,e^p;Q)=0.
\eeq
Moreover, as discussed in the previous sections, this
classical geometry of the mirror captures all the disc corrections involving worldsheets ending on this special Lagrangian.

On the other hand, using geometric transition of the large $N$ Chern-Simons theory, the exact open topological string partition function of the brane on $L_K$, in the resolved conifold, is equivalent to the computation of HOMFLY invariants (or knot homologies which arises in the Nekrasov deformation of topological strings and refined Cherns-Simons theory \cite{MS, MS2}).
The exact wave function of the brane, computed by Chern-Simons theory and the topological string, is 
\beq\label{exacpsi}
 \Psi_{K}(x, g_s, N)= \sum_n H_n(K) \; e^{-nx}.
\eeq
The  large $N$ limit of the amplitude is equivalent to summing over all planar diagrams of the Chern-Simons theory.  This is dominated by the disk instanton contribution that sums up all the planar graphs in Chern-Simons theory with one boundary on the D-brane. This can be written as
\beq\label{semi}
\Psi_{K}(x, g_s, Q) \sim exp{({1\over g_s} \int p_K(x,Q) dx}  ).
\eeq
Mirror symmetry and large $N$ duality now imply that \cite{OV, LMV}
$$
p=p_K(x,Q)={\rm lim}_{g_s\rightarrow 0}\sum_{n=0}^{\infty} g_s\langle tr_K U^n \rangle e^{-nx}
$$
lives on the mirror curve \eqref{mirror}. Conversely, from HOMFLY polynomial $H_{R_n}(q,N)$ colored by symmetric representations, and extracting the classical piece we can read off the curve of the mirror Calabi-Yau manifold. We will give examples of this in section 6.

\subsection{Chern-Simons and The Quantum Mirror}
Before we go on, note that instead of simply studying the classical equation of the mirror, Chern-Simons theory also gives us the exact quantum mirror geometry. 
In \cite{IH} it was found that the branes on the mirror Riemann surface are annihilated by a quantum operator that quantizes the classical geometry.
However, except when the underlying Riemann surface has genus zero, finding the exact brane wave functions is hard by direct methods, and correspondingly the quantum mirrors
are not known in general. Large $N$ duality and Chern-Simons theory gives us a way to find the exact brane wave function of the mirror. Thus, from the colored HOMFLY polynomial, without taking any limits, we can derive the quantum mirror Riemann surface. 

The classical approximation to the wave function \eqref{semi} determines the classical mirror geometry
$$
F_K(e^x,e^p, Q)=0.
$$
As discussed in \cite{IH}  this should be viewed as the classical $g_s\rightarrow 0$ limit of the operator equation
\beq\label{quantmirr}
F_K(e^{\hat x},e^{\hat p}, Q)\Psi_K(x)=0
\eeq
where we view ${\hat x}$ as a multiplication by $x$, and $\hat{p} = g_s \partial_x$, so that
$$
[{\hat p}, {\hat x}] = g_s.
$$
In this way Chern-Simons theory on the $S^3$ gives us a way to solve the topological string on the mirror of the brane and get the
quantum Riemann surface associated to the knot. This is an example of quantum Riemann surfaces that emerge from studying D-branes in the topological string \cite{IH}.

Before we go on, note that we can view $\Psi_K(x,Q)$ as a discrete Fourier transform of $H_{R_n}(K,q,Q)$. Correspondingly, an equivalent way to represent the quantum mirror curve is via its action on 
$H_{R_n}(K)$. This gives
\beq\label{quantmirr}
{\hat F}_K(e^{\hat x},e^{\hat p}, Q)H_{R_n}(K)=0
\eeq
where now 
$$
e^{\hat p} H_{R_n}(K) = q^n H_{R_n}(K) , \qquad
e^{\hat x} H_{R_n}(K)  =  H_{R_{n+1}}(K) .
$$

\section{Relation to the A-polynomial and the Volume Conjecture}

We have seen in the previous sections that mirror symmetry and large $N$ duality associate a quantum Riemann surface of the form
\beq
{\hat F}_K(e^{\hat x},e^{\hat p};Q)=0.
\eeq
to a knot in $S^3$. Quantum Riemann surfaces of this type have appeared in the literature in the context of the quantum volume and AJ conjectures. We will now explain what the connection is.

Consider Chern-Simons theory with gauge group $G=SU(2)$ and Chern-Simons level $k$ on $S^3$ with a knot $K$ inside it.  The colored Jones polynomial $J_n(K)$ computes the expectation value of a Wilson loop along the knot, in the spin $n/2$  representation of $G$.
It was noticed \cite{garo,Gukov:2003na} that the colored Jones polynomial $J_n(K)$ satisfies a difference equation,
\beq\label{AJ}
\hat{A}_K({\hat x}, {\hat p}) \; J_n(K) =0,
\eeq
where ${\hat x}$ and ${\hat p}$ are defined by
$$
e^{\hat x} J_n= J_{n+1}, \qquad
e^{\hat p} J_n = q^n J_{n},
$$
where $q = e^{\frac{2 \pi i}{ k+2}} = e^{\hbar}$, where $\hbar$ is the effective coupling constant of Chern-Simons theory.\footnote{In \cite{garo}, one used the normalization in which the Jones polynomial of the unknot $J_n(\bigcirc) = 1$ for all $n$. We will not do that.} 
In particular,
\beq\label{con}
[{\hat x},{\hat p}] = \hbar.
\eeq
Notice that $e^{\hat p} J_n = q^n J_{n}$
implies that $p$ can be identified with $ng_s$.  In fact this is most often
how the wave function $\tilde \psi(p)$ of the $N=2$ theory has been defined in the context
of the volume conjecture.  Namely by the analytic
continuation of
$$\tilde \Psi(p)=J_{p/g_s}(K)   $$
In the classical $\hbar \rightarrow 0$ limit $A_K$ becomes a plane holomorphic curve,
$$
A_K(x,p)=0.
$$ 
The curve that arises in this way is called the characteristic variety of the knot. Moreover, they are closely related to another Riemann surface from classical mathematics,
the A-polynomial  is defined by studying the moduli space of flat $SL(2,{\bf C})$ connections on the complement of the knot $K$ in the $S^3$. It was studied in \cite{Apol}. Given a knot $K$ in $S^3$ one considers its complement,
$$M_K=S^3/K
$$ 
Cutting out the neighborhood of the knot form an $S^3$, one obtains a three manifold with a $T^2$ boundary.  
The the coordinates $p$ and $x$ parameterize the flat $SL(2,{\bf C})$ connections,
around the cycles $\mu$ and $\lambda$ that generate $\pi_1(T^2)$. Here $\mu$ is the meridian that links the knot, and the $\lambda$ is the longitudinal cycle that runs parallel to the knot. The condition that these extend smoothly from the boundary $T^2$ to the interior of $M_K$ imposes a relation between $x$ and $p$ that is the A-polynomial $A_K(x,p)=0$.
  We now explain the relation to the Riemann surfaces $H_K(x,p,Q)$ that arise in the context of topological strings.

The colored Jones polynomial is a special case of the HOMFLY polynomial colored by symmetric representations, where we restrict to $N=2$ and set $Q=q^N=q^2$,
$$
J_n(K, q) = H_{R_n}(K,q^2,q).
$$
This immediately implies that the quantum A-polynomial $A_K({\hat x},{\hat p})$ arises as a special case of the quantum mirror curve where we would replace
$Q=q^2$.  This identifies the corresponding quantum Riemann surfaces.
The main difference between the descriptions of the wave functions is that in the
context of the topological string the $x$ coordinate, which is holonomy along longitudinal
cycle of the knot is the natural variable, whereas
in the context of the volume conjecture the $p$ coordinate, which is the holonomy along
the meridian is more natural.  This in particular means that if
we consider the Fourier transform of the wave function we have studied in the context
of topological strings we would get the wave function relevant for the quantum volume or AJ conjecture:
\begin{align*}
{\tilde \Psi}_K(p)&=\int dx \ {\rm exp}({px\over g_s}) \ \Psi_K(x) \\
&=\int dx\ {\rm exp}({px\over g_s}) \sum_n {\rm exp}(-nx) J_n(K)=J_{p/g_s}(K)
\end{align*}
Moreover we expect that as we take $q$ and $Q$ to $1$, we obtain the classical A-polynomial. 
Furthermore, the explanation of why the classical A-polynomial is related to Jones polynomial
and how it relates to the volume of the hyperbolic knot complement now follows because
all are related to $SL(2)$ Chern-Simons as was already noted in \cite{Gukov:2003na}.
In particular in the classical limit fixing the $x,p$ holonomies and solving the $SL(2)$
Chern-Simons equations, lead to flat solutions and thus they should lead to the constraint
captured by the A-polynomial.
There is a subtlety here, as it turns out that the limit $q\rightarrow 1$ leads to a polynomial
which is not exactly the A-polynomial but has the A-polynomial as a factor.
We can offer an explanation of this, namely these could be interpreted as different
branches of the brane geometry where the Lagrangian brane $L_K$ has different reconnections
with branes on $S^3$ as well as possibly moved off of $S^3$. 
For example, for the unknot we get in this limit $(1-e^x)(1-e^p)=0$.  One solution $e^x=1$, where $x$ corresponds to longitudinal direction along $K$, 
 is part of the classical A-polynomial and signifies
the branch where the brane $L_K$ has reconnected with one of the branes on $S^3$ and has
trivialized the holonomy around the non-trivial circle in $L_K$. 
The other branch $e^p=1$ could signify the geometry where the $L_K$ is lifted
off of $S^3$, in which case $e^p$ is trival because it corresponds to the contractible
circle in $T^2\subset L_K$.  It would be interesting
to study this multi-branch structure for more general knots, and predict
the additional factors that multiply the classical A-polynomial.  

Considering the limit of $q\rightarrow 1$ keeping $Q$ fixed,
leads to a deformation of the classical A-polynomial to a Q-deformed version.  More precisely, we have the following sequence of limits:
$$\hat{F}_K(e^{\hat x},e^{\hat p};Q){\rightarrow}_{q\rightarrow 1} F_K(e^x,e^p,Q) \rightarrow_{Q\rightarrow 1}A_K(e^x,e^p)$$
and
$$\hat{F}_K(e^{\hat x},e^{\hat p};Q)\rightarrow_{Q\rightarrow q^2} \hat{A}_K(e^{\hat x},e^{\hat p}) \rightarrow_{q\rightarrow 1}A_K(e^x,e^p)$$
More precisely in the last identity we expect that $A_K(e^x,e^p)$ is a factor of $F_K(e^x,e^p;1)$.
Thus, 
it simply follows from large $N$ duality arguments
and mirror symmetry that the mirror curve associated to the knot $K$, $F_K(e^x,e^p,Q)$ is a Q-deformation of a polynomial which has the classical A-polynomial of the knot $A_K(e^x,e^p)$ as a factor.
This is the same as the classical mirror geometry
associated to $K$.
In other words, it simply follows from large $N$ duality arguments
and mirror symmetry that there exists a Q-deformation of a polynomial $A_K(e^x,e^p)$  which has
the classical A-polynomial as a factor, which in turn is the same as the classical mirror geometry
associated to $K$.

We can already make a few checks about this statement.  In particular let us consider the limit $Q\rightarrow 1$.  In this
case we should obtain the mirror to the conifold at the singular point, where the size of the ${\bf P}^1$
is zero.  If $A_K(e^x,e^p)$ is in fact the mirror geometry for the conifold, then it better be true that
the periods we compute for $A$ using the mirror 1-form $pdx$ are zero.  More precisely the period
should be zero, up to the ambiguity of adding an integer multiple of $2\pi i$, because $Q=1$ only
leads to $t=2\pi i n$ for some $n$.  Indeed this is one of the deep properties\footnote{
More precisely, they are known to have rational periods, which we interpret as defining
the correct cover of the variables to lead to integral periods.} of the classical
A-polynomial for knots \cite{boyd}! This is already a first indication of the power of our conjecture, which
automatically provides an alternative explanation of this fact.
For general $Q$,  the mirror curve $A_K(Q)=0$ should have the same 
$log(Q)$-periods as the canonical mirror of the conifold.  We will show this in the example of $(m,n)$ torus knot curves, for arbitrary $(m,n)$.  Unlike the A-polynomials, the $A_K(Q)$ polynomials of these curves are highly non-trivial, as we will see. Hence this is a strong test of our conjecture.

\section{Examples}

As we discussed above, the canonical of mirror of $X^v={\cal O}(-1)\oplus {\cal O}(-1) \rightarrow {\bf P}^1$
is
$$
uv = (1-e^p)-e^x(1-Qe^p).
$$
In this section, we will explain that the canonical mirror is associated to the simplest knot in the $S^3$, the unknot. Then, by changing the knot type, we will give examples of different mirrors that can arise. For each of these, we conjecture that topological strings on them are indistinguishable.

\subsection{The Unknot}

The canonnical mirror Riemann surface,
$$
 F(e^x,e^p,Q) = (1-e^p)-e^x(1-Qe^p) =0
$$
 is a quantum corrected moduli space of a simple Lagrangian brane $L_O={\bf R}^2\times S^1$ in $X={\cal O}(-1)\oplus {\cal O}(-1)\rightarrow {\bf P}^1$, where the $S^1$ can be roughly thought of as wrapping around the equator of the ${\bf P}^1$, and where the $\bf{R}^2$ is the Lagrangian subspace of the ${\cal O}(-1)\oplus{\cal O}(-1)$ fiber to the ${\bf P}^1$. From our previous discussion it is natural to ask which knot in the $S^3$ is this associated to. As explained in \cite{OV}, the Lagrangian brane $L_O$ in $X^v$ corresponds to the unknot in the $S^3$. The mirror geometry could have been derived from $G=SU(N)$ Chern-Simons theory along the lines we described above. Recall that for the unknot, the HOMFLY polynomial colored by the $n$'th symmetric representation $R_n$ equals
$$
H_{n}(\bigcirc) =  S_{0R_n}= \frac{[N][N+1] \ldots[N+n-1]}{[1][2]\ldots [n]} 
$$
where $S_{RQ}$ is the $S$ matrix of $SU(N)_k$ Chern-Simons theory, and we need its element with $Q=0$, $R=R_n$.
We can rewrite this, in terms of $Q=q^N$ as follows:
\beq\label{unknot}
H_{n}(\bigcirc) = Q^{-{n\over 2}}q^{n\over 2}\frac{(1-Q)(1-Qq) \ldots(1-Q q^{n-1})}{(1-q)(1-q^2)\ldots (1-q^n)} 
\eeq
where $[n] = q^{n/2} - q^{-n/2}$. From the recurrence relation satisfied by $H_n$ 
$$
(1-q^n) H_{n} - Q^{-{1\over 2}} q^{{1\over 2}} (1-Q q^{n-1}) H_{n-1} =0.
$$
we can read off the equation of the quantum mirror. Up to a redefinition of the holonomy variable $x$ by a constant shift,
we get
$$
{\hat F}_{\bigcirc} (e^{\hat x},e^{\hat p},Q)= 1-e^{\hat p} - e^{\hat x} (1-Q e^{\hat p}) =0
$$ 
acting on $H_n $, or equivalently, on the wave function
$$
\Psi_O(x) = \sum_{n=0} H_n e^{-nx}.
$$
In particular, in the classical limit, ${\hat F}_{\bigcirc} (e^{ x},e^{p},Q)$ gives the Riemann surface of the cannonical mirror Calabi-Yau $Y$.

\subsection{Torus knots}

For any knot $K$ we expect, via large $N$ duality and mirror symmetry, to obtain a mirror Calabi-Yau to $X$. A large family of examples is provided by torus knots.

The colored $SU(N)_k$ polynomial for an $(m,n)$ torus knot is given by Rosso-Jones formula \cite{RJ}.
It will be useful for us to recall a derivation of it, as understanding it will allow us to make some shortcuts later on. If we view the $S^3$ as a $T^2$ fibration over an interval, we can take an $(m,n)$ torus knot to wind around the $(m,n)$ cycle of the $T^2$. This provides a way to compute the corresponding knot invariants: we think about the holonomy 
$$  {\rm Tr}_R \,U_{(m,n)}  =  {\rm Tr}_R \,e^{m x+n p}
$$
as  the holonomy of an unknot knot winding $m$ times around the $(1,0)$ cycle, but with fractional framing $f=m/n$, corresponding to the shift of $x$ to $x+m/np$. 
$$
\langle {\rm Tr}_R  \,e^{m x + n p}\rangle  = \langle {\rm Tr}_R  \,e^{n (x + m/n p)}\rangle
$$
Using the fact that an
unknot in framing $f$, has expectation value $\langle Tr_Q e^{x + f p}\rangle = S_{0Q} T_Q^f$, 
where $S$ and $T$ are the $S$ and the $T$ matrices of $SU(N)_k$ Chern-Simons theory,
and expanding $ {\rm Tr}_R U^n$ in terms of ${\rm Tr}_Q U$
$${\rm Tr}_R \, U^n = \sum_{Q}  C^{(n)}_{R Q} {\rm Tr}_Q \, U.
$$
we get the Rosso-Jones formula.
\beq\label{RJ}
H_R^{(m,n)} =  \langle {\rm Tr}_R  \,U_{(m,n)} \rangle = \sum_P  \, C^{(n)}_{R P}  \, T_P^{m/n}  \, S_{P0}.
\eeq 
To find the D-brane moduli space, we can proceed in two equivalent ways. We can use the exact HOMFLY
to find the quantum operator ${\hat F}_{(m,n)}(e^{\hat x},e^ {\hat p},Q)$ annihilating the brane wave function
$$
{\hat F}_{(m,n)}(e^{\hat x}, e^{\hat p},Q)\Psi(x)=0
$$
and then take the classical $g_s$ to zero limit to find the equation of the mirror. However, this is an overkill, if all we want is the classical geometry of the mirror. Moreover, should one want to find the exact recurrence, one would have to rewrite the colored HOMFLY in some way, to have a more compact formula for the $n$-th term of the sum. It would be very interesting to obtain such a formula, analogously to what was done for the figure 8 knot in \cite{kawa} and recently in \cite{morozov}\footnote{As we prepared this paper for publication, \cite{GS2} appeared, where just such a formula was found, and in fact extended to the refined setting. Using the formulas in \cite{GS2}, one could in fact obtain also the quantum mirror curves.}. Instead, one can simply use the leading large $N$ limit of the HOMFLY amplitude, which was obtained in \cite{marino}.\footnote{In \cite{marino} the authors studied mirrors of torus knots.
The curves obtained there are different than the ones we find. The difference, roughly, is that in their case, the mirror brane corresponds not to a single point on the Riemann surface as is the case for us, but to $n$ points moving together where $n$ is determined by the $(m,n)$ type of the torus knot. The authors showed that one can reproduce the $g_s$ corrections to some closed and open string amplitudes of the conifold with the branes corresponding to this knot. It would be nice to elucidate the physical interpretation of the curves obtained in that paper, as one is sure to exist.}
 For the $(m,n)$ torus knot, one finds
$$
\Psi(x)\sim \exp({1\over g_s} \int p(x)dx)
$$
where 
\beq\label{torp}
p(x) = \sum_{k=0}^{\infty} {V}_{k}^{(m,n)} e^{-kx} 
\eeq
and \cite{MarinoVafa, marino}
\beq\label{vev}
V_{k}^{(m,n)} = \frac{1}{(kn)!} \sum_{j=0}^{kn} {kn\choose j} (-1)^{kn+j} Q^{j + km-kn} \prod_{i=-j+1}^{kn-1-j}\bigl(km  -i\bigr)
\eeq
While these expressions do not have a manifest symmetry corresponding to exchanging $(m,n)$, once the sum in  \eqref{vev} is performed, the symmetry is restored.  As we explained in section 4, the coefficients $V_{k}^{(m,n)}$ have a direct Chern-Simons interpretation, as a limit of the expectation value of the Wilson loop along the knot
$$
V_{k}^{(m,n)} = {\lim}_{g_s\rightarrow 0} \; \langle {\rm tr} U^k_{(m,n)}\rangle g_s .
$$
From \eqref{torp} and \eqref{vev} for specific $m,n$ we can write down the mirror curves this gives rise to, by simply asking for
$(x,p(x))$ to lie on the curve. We will explain later the method we used to obtain the curves, but for now it suffices to give examples
of the results one gets.

\subsubsection{The $(2,3)$ knot}

The simplest non-trivial torus knot is the $(2,3)$ knot, the trefoil. For the trefoil, we find that $(x,p(x))$ lie on the following curve:
$$
F^{(2,3)}(\alpha,\beta,Q)=
(1-Q \beta )+\left(\beta ^3-\beta ^4+2 \beta ^5-2 Q \beta ^5-Q \beta ^6+Q^2 \beta ^7\right) \alpha +\left(-\beta ^9+\beta ^{10}\right) \alpha ^2
$$
For legibility of the formulas, in this and the following examples, we will denote
$$e^x=\alpha, \qquad e^p=\beta.
$$
To be precise, to reproduce the leading large $N$ limit of the colored HOMFLY polynomial, we need to replace $\alpha$ by $\alpha Q^{5/2}$. While fractional powers of $Q$ are natural in knot theory, they are not natural in the topological string, so we prefer to present the curve in this way. 
At $Q=1$ limit, this becomes 
$$
F^{(2,3)}(\alpha,\beta,1)=(-1+\beta ) \left(1+\alpha  \beta ^3\right) \left(-1+\alpha  \beta ^6\right)
$$
The classical $A$ polynomial of the trefoil knot,
$$
A_{(2,3)}(\alpha,\beta)=\Bigl(1+\alpha\Bigr) \Bigl(-1+\alpha\beta^3\Bigr) 
$$
corresponds to the last two factors in $F_{(2,3)}(\alpha, \beta, 1)$, after the change of framing that takes $\alpha$ to $\alpha \beta^{-3}$. In section 5 we suggested the interpretation of the remaining factor, $(-1+\beta)$, as corresponding to the branch of the moduli space where the branes on $L_K$ move off
the $S^3$. On this branch, the corresponding cycle becomes contractible in $L_K$, so the holonomy around it vanishes, setting $\beta$ to $1$.

\subsubsection{The $(2,5)$ knot}
For the $(2,5)$ knot, we get the following curve.
\begin{align*}
F^{(2,5)}(\alpha,\beta,Q)&=
(1-Q \beta )+(2 \beta ^5-\beta ^6-Q \beta ^6+3 \beta ^7-4 Q \beta ^7+Q^2 \beta ^7-2 Q \beta ^8+2 Q^2 \beta ^8\\
&+2 Q^2 \beta ^9-2 Q^3 \beta ^9-Q^3 \beta ^{10}+Q^4 \beta ^{11}) \alpha +(\beta ^{10}-\beta ^{11}+2 \beta ^{12}\\
&-2 Q \beta ^{12}-2 \beta ^{13}+2 Q \beta ^{13}+3 \beta ^{14}-4 Q \beta ^{14}+Q^2 \beta ^{14}-Q \beta ^{15}-Q^2 \beta ^{15}\\
&+2 Q^2 \beta ^{16}) \alpha ^2+(-\beta ^{20}+\beta ^{21}) \alpha ^3
\end{align*}

This reproduces the HOMFLY, up to a shift of $\alpha$ by $Q^{7/2}$. In the $Q=1$ limit, the curve degenerates to
$$
F^{(2,5)}(\alpha,\beta,1)=(-1+\beta ) \left(1+\alpha  \beta ^5\right)^2 \left(-1+\alpha  \beta ^{10}\right)
$$
The classical A-polynomial is a factor in $F^{(2,5)}(\alpha,\beta,1)$, after one changes the framing by replacing $\alpha \rightarrow \alpha \beta^{-5}$.

\subsubsection{The $(2,7)$ knot}
For the $(2,7)$ knot, we get the following curve.
\begin{align*}
F^{(2,7)}(\alpha,\beta,Q)&=
(1-Q \beta )+(3 \beta ^7-\beta ^8-2 Q \beta ^8+4 \beta ^9-6 Q \beta ^9+2 Q^2 \beta ^9-3 Q \beta ^{10}\\
&
+4 Q^2 \beta ^{10}-Q^3 \beta ^{10}+3 Q^2 \beta ^{11}-4 Q^3 \beta ^{11}+Q^4 \beta ^{11}-2 Q^3 \beta ^{12}+2 Q^4 \beta ^{12}\\
&
+2 Q^4 \beta ^{13}-2 Q^5 \beta ^{13}-Q^5 \beta ^{14}+Q^6 \beta ^{15}) \alpha +(3 \beta ^{14}-2 \beta ^{15}-Q \beta ^{15}+6 \beta ^{16}\\
&
-8 Q \beta ^{16}+2 Q^2 \beta ^{16}-3 \beta ^{17}+2 Q \beta ^{17}+Q^2 \beta ^{17}+6 \beta ^{18}-12 Q \beta ^{18}+10 Q^2 \beta ^{18}\\
&
-4 Q^3 \beta ^{18}-3 Q \beta ^{19}+2 Q^2 \beta ^{19}+Q^3 \beta ^{19}+6 Q^2 \beta ^{20}-8 Q^3 \beta ^{20}+2 Q^4 \beta ^{20}\\
&
-2 Q^3 \beta ^{21}-Q^4 \beta ^{21}+3 Q^4 \beta ^{22}) \alpha ^2+(\beta ^{21}-\beta ^{22}+2 \beta ^{23}-2 Q \beta ^{23}-2 \beta ^{24}\\
&
+2 Q \beta ^{24}+3 \beta ^{25}-4 Q \beta ^{25}+Q^2 \beta ^{25}-3 \beta ^{26}+4 Q \beta ^{26}-Q^2 \beta ^{26}+4 \beta ^{27}-6 Q \beta ^{27}\\
&
+2 Q^2 \beta ^{27}-Q \beta ^{28}-2 Q^2 \beta ^{28}+3 Q^2 \beta ^{29}) \alpha ^3+(-\beta ^{35}+\beta ^{36}) \alpha ^4
\end{align*}

In the $Q=1$ limit, we get
$$
F^{(2,7)}(\alpha,\beta,1)=(-1+\beta ) \left(1+\alpha  \beta ^7\right)^3 \left(-1+\alpha  \beta ^{14}\right)
$$
The classical A-polynomial is a factor in $F^{(2,7)}(\alpha,\beta,1)$, after one changes the framing by replacing $\alpha \rightarrow \alpha \beta^{-7}$.

\subsection{Periods of the Torus Knot Curves}

Our claim is that for every knot, in particular for every $(m,n)$ torus knot, the curve gives rise to a mirror Calabi-Yau 
$$
uv = F^{(m,n)}(e^x,e^p,Q)
$$
of the resolved conifold. From the SYZ perspective, all the mirrors are on the same footing, as there is no a-priori reason to favor one curve over the other. Note that the curves one gets are very high genus. The $F^{(2,3)}$ is nominally a genus three curve, while $F^{(2,5)}$ is a genus ten curve. Yet, we expect the topological strings on them to be equivalent. In particular, the classical periods of these cures should be equivalent. We will now prove that in fact for $all$ torus knots, the latter statement holds. Moreover, the methods we will use to prove this will also be closely related to how we found the curves above, so we will have an opportunity to explain this.

To show that the periods of 
$$
\int p(x) dx
$$
on a curve
$$F^{(m,n)}(e^x,e^p,Q)=0
$$
are the same as the periods of the conifold, we will find suitable variables in terms of which we can rewrite as 
$$
\int p(x)dx = \frac{1}{ n} \sum_{k=0}^{n-1} \int p_k(x_k)dx_k  
$$
where each of the one forms $\int p_k(x_k)dx_k $ in fact has the same periods, and indeed the same periods as on the mirror of the conifold.
This, together with the $1/n$ factor out front, will imply that $\int p(x)dx$ itself has the same periods, as the one form on the canonical mirror, thus proving our claim. To see how such variables may arise we need to back up, all the way to the beginning and the derivation of the Rosso-Jones formula.

In the presence of a Lagrangian brane corresponding to an $(m,n)$ torus knot, the topological string computes the expectation value of
$$
\Psi(e^x)=\langle det^{-1}(1\otimes 1 - U^n \otimes e^{-x})\rangle  
$$
where $U = e^{x+ m/n p}$ corresponds to an unknot in the fractional framing. We can write this as
$$
\Psi(e^x)=\langle\prod_{\ell=1}^n  det^{-1}(1\otimes 1 - \omega^ \ell U \otimes e^{-x/n})\rangle
$$
where $\omega$ is a primitive, $n$-th root of unity. As long as we are interested in the semi classical limit (or more precisely, as long as we are interested in single trace contributions to the amplitude where the multi-trace contributions are suppressed relative to this by powers of $g_s$), we can think of this as a wave function of $n$ particles, evenly distributed
$$
\Psi(e^x)\sim \prod_{\ell=1}^n \langle det^{-1}(1\otimes 1 - \omega^ \ell U \otimes e^{-x/n})\rangle = \prod_{\ell=1}^n \Psi_*(e^{-x/n} \omega^{\ell})
$$
Each of the $n$ particles simply corresponds to an unknot in fractional framing $f=m/n$. If we denote their coordinates and momenta as $(x_{\ell}, p_{\ell})$, ${\ell}=1, \ldots n$ each of these lives on the Riemann surface which is a simple mirror of the conifold, but with $x$ shifted by $(m/n)p$, corresponding to the change of framing
$$
e^{m{p}_{\ell}/n} (1-e^{p_{\ell}}) -e^{-x_{\ell}} (1-e^{ p_{\ell}} Q)=0.
$$
The closed periods are independent of the change of framing \cite{AKV} so in fact each of the one forms $\int p(x_{\ell})dx_{\ell}$ has the same periods as the conifold. Now in order to relate this to $\Psi(x)$ we need to identify
$e^{x_{\ell}} =e^{x/n} \omega^{\ell}$. This identification is not in $SL(2,{\bf Z})$ -- it rescales $pdx$ by $1/n$, so in fact each particle has the same periods as the canonical mirror of the conifold, albeit rescaled by $1/n$. But, this is exactly what we need to recover the periods of the mirror.
In summary, while $(p(x),x)$, for a generic $(m,n)$ torus knot live on a complicated curve, there is a way to introduce $n$ sets of variables $(p_\ell, x_{\ell})$ in terms of which it becomes apparent that one is in fact studying the conifold.

This idea also helps in finding the explicit curve on which $p$ and $x$ live, $F^{(m,n)}(p,x).$ Namely, as long as we are interested  in obtaining the classical curve only, we are allowed to replace $\Psi(e^x) $ by $\prod_{\ell=1}^n\Psi_{\ell}(e^{{x}_{\ell}})$. The latter has the form of a q-hypergeometric multi-sum. There are by now very powerful computer programs that can be used to find the corresponding $q-$ recurrence relations. Taking the classical limit of this, gives us the curves we need.
Explicitly, we have
%
%
%
%
the brane amplitude of the framed unknot:
$$
\Psi^{(f)}(x) = \sum_{R_k} T_{R_k}^{f} S_{R_k0} e^{-kx}= \sum_{k=0} H_k(\bigcirc) q^{-f k^2/2} q^{f k(N-1)/2}e^{-kx}
$$
where $H_k(\bigcirc)$ is the HOMFLY polynomial of the unknot \eqref{unknot}. Taking $f=m/n$ gives $\Psi_*(x)$ we used,  for a 1-particle wave function in our $n$-particle system.
%
%
 %
%
\subsection{Figure Eight Knot}

The figure eight knot HOMFLY polynomial, colored by symmetric representation with $n$ boxes is given by
\begin{align*}
H^{\bf 4_1}_n=&\sum_{i=0}^n\sum_{j=0}^i (-1)^{i} Q^{3+2 i- 3n\over 2} q^{-2 n^2+i^2+4 i n+2j+7n-5i-5\over 2}\\
&\times {[\left(q^{-i},q\right)_j]^2[\left(q^{1-n},q\right)_i]^2\left(Qq^{-1- i+j+ n},q\right)_{i-j}\left(Q,q\right)_{n-1}\over
 \left(q,q\right)_i\left(q,q\right)_j \left(q,q\right)_{n-1}  \left(q^{n- i+j},q\right)_{i-j}}
\end{align*}
This result was provided to us by K. Kawagoe \cite{kawa}, based on unpublished work.
(see also the related work \cite{kawavol}). Recently, another expression for the colored HOMFLY in these representations was obtained in \cite{morozov}. The formulas should be equivalent, although they are not manifestly so.  In the above, $(a,q)_n=\prod_{k=0}^{n-1}(1-aq^k)$ is a version of $q$-factorial. The knot invariant is in the form of $q-$hypergeometric multi-sum. Recurrences of such sums can be found by a powerful program for finding recurrences of such sums, "HolonomicFunctions.m"\footnote{The recurrence corresponding to this sum was computed by Christoph Koutschan, who also wrote the program. We are indebted to him for patiently explaining how to use his program.  The program is publicly available from\\
http://www.risc.jku.at/research/combinat/software/HolonomicFunctions/index.php}.

From this, we can find the mirror geometry corresponding to the figure eight knot.
\begin{align*}
F_{{\bf 4}_1}(\alpha,\beta,Q)&=
(\beta ^2-Q \beta ^3)+(-1+2 \beta -2 Q^2 \beta ^4+Q^2 \beta ^5) \alpha \\&+(1-2 Q \beta +2 Q^2 \beta ^4-Q^3 \beta ^5) \alpha ^2+(-Q^2 \beta ^2+Q^2 \beta ^3) \alpha ^3
\end{align*}
This is obtained as the $q\rightarrow 1$ limit of the quantum ${\hat F}_{{\bf 4}_1}$ polynomial. Viewed as a limit of the quantum curve, one
finds additional trivial pieces, 
$(-1 + \beta^2)^2 (\alpha - Q)^2$ terms. At $Q=1$, our curve becomes
$$
F_{{\bf 4}_1}(\alpha,\beta,1)=(-1+\alpha ) (-1+\beta ) \left(\alpha -\alpha  \beta -\beta ^2-2 \alpha  \beta ^2-\alpha ^2 \beta ^2-\alpha  \beta ^3+\alpha  \beta ^4\right)
$$
This agrees with the A-polynomial of the figure eight knot, up to a factor $\left(-1+\beta\right)$.
\section{Relation with Knot Contact Homology}
After the completion of this work, and during preparation of this manuscript, we became
aware\footnote{We thank Marc Culler, Craig Hodgson and Henry Segerman for pointing
this out.} of the work by Ng \cite{Ng10} and collaborators \cite{nget}, which seems
to be related with the present work.  For a readable review
of the basic ideas see \cite{Ng04} and references therein. Their work uses symplectic field theory and knot contact homology. The relation of these with topological strings, Gromov-Witten theory and Chern-Simons theory was not observed previously. 
Below we propose an explanation of such a relation.

  Both this paper and the ones by Ng and collaborators
use the construction in \cite{OV} associating to each knot a Lagrangian brane.
In the work \cite{Ng10} one starts with the Lagrangian brane in the $T^*S^3$ but
deletes the zero section of the $S^3$.  In other, words, one is focuing on the
geometry far from the apex of the cone where the geometry is $S^3\times S^2 \times {\bf R}$ which shares the basic feature of the geometry
{\it after} the transition, where $S^2$ is no longer contractible.  Furthermore one considers
the $L_K$, i.e. the Lagrangian $L_K$ associated with the knot, which has the topology of
$T^2\times {\bf R}$.  It is natural to view ${\bf R}$ as the `time' direction.
  Here $T^2$ can be viewed as the geometry of the Lagrangian
brane at infinity.  The knot leaves its imprint on how $T^2$ is embedded in
$S^3\times S^2$.  Then one considers the physical open string states ending on $T^2$.
These are in 1-1 correspondence with `Reeb chords'  $a_i$ which are open string trajectories
beginning and ending on $T^2$ and which extend along Reeb flows (which correspond
to complex pairing of ${\bf R}$ with a direction on $S^3\times S^2$, and so the holomorphic
maps can end on them).   In the physical setup we would say that the $a_i$ are classically annihilated by the
BRST symmetry $Q_B$:
$$Q_B\cdot a_i=0$$
However disk instantons could modify the operation of $Q_B$.  This is similar to Witten's formulation
of Morse theory \cite{Witten:1982im} where the critical points of the Morse function
are in 1-1 correspondence with vacua, but instanton corrections, which correspond
to gradient flows, modify the supersymmetry algebra.  In the case at hand the role
of the gradient flows are played by
disc instantons.    Consider one such disc $D$ which at $t\rightarrow +\infty$ starts with the Reeb chord $a_i$ and at $t\rightarrow -\infty$ approaches
the Reeb chords $a_{ij}$ as ordered multi-pronged strips with $j=1,...,k_D$. Let the boundary of the disc
map to $H_1(T^2)$ (upon choosing some fixed intervals capping the ends of the Reeb
chords), which picks up the holonomy factor ${\rm exp}(n^D_iu+m^D_iv)$ from the Wilson lines on the Lagrangian
brane.  Let $n_D$ denote
the intersection of this disc with the 4-cycle dual to $S^2$, i.e. ${\bf R}\times S^3$.   Then
Ng et.al. find a deformed $Q_B$ operator (which we interpret as the quantum corrected $Q_B$)  given by
$$Q_B\cdot a_i=\sum_{disc \ D} Q^{n_D} e^{n^D_iu+m^D_iv} a_{i_1}...a_{i_{k_D}}.$$
Thus the open string states correspond
to non-trivial elements of the deformed $Q_B$ cohomology.  The existence of
a 1-dimensional representations of this cohomology leads to a constraint $A(e^u,e^v,Q)=0$.
Such one dimensional representations can be interpreted as the disc corrected moduli space of one such Lagrangian brane after the large $N$ transition.  Using the large $N$ duality this should be 
 the same as the Q-deformed A-polynomial we defined, which was the quantum corrected moduli space of the single brane $L_K$.  We computed this {\it before} the transition using
the color HOMFLY polynomial at leading order in large $N$.
We have checked that for the examples $(2,3),(2,5)$ torus knots and the figure-8 knot,  for which the knot contact homology has been computed are in accord
with computations done here\footnote{We thank Lenhard Ng for communicating
his results.}.  This agreement is thus a consequence of the large $N$ duality.

\section{Concluding Remarks}

The embedding of Chern-Simons theory in string theory and its large $N$ duality
lead to the statement that the knot invariants of the HOMFLY type
as well as their  Khovanov-type refinements can be deduced in principle
once we know the Lagrangian brane associated with the knot on the large $N$ dual
geometry.  In this paper, we have shown that, using a generalized SYZ conjecture
adapted to the local Calabi-Yau geometries, to every knot we can associate a mirror Calabi-Yau $Y_K$.
This mirror is captured by a classical polynomial $A_K(x,p;Q)$ which as $Q\rightarrow 1$
has the classical A-polynomial of the knot as a factor.  Since HOMFLY and their refinement
in the form of Khovanov invariants and knot homologies can be computed perturbatively
starting with the classical B-model geometry of $Y_K$, it implies that the classical polynomial $A_K(x,p;Q)$ is at least
as refined as the knot homologies.

Namely, having related a knot $K$ to the Calabi-Yau $Y_K$, we relate quantum and homological invariants of a knot $K$ to (open) topological string amplitudes on $Y_K$ and their refinement.  This implies that, to gain any new information about distinguishing knots by going to $SL(N)$ or Khovanov type invariants, we would need to get new information about distinguishing Calabi-Yau manifolds, beyond that captured by classical geometry, by considering Gromov-Witten or topological string amplitudes on $Y_K$ instead. But, we do not:  at least in principle, Gromov-Witten theory is a machine that uniquely assigns to a classical geometry, quantum amplitudes. 
We thus expect $A_K(x,y;Q)$ to completely capture all HOMFLY knot
invariants, including its refinements.
The one possible pitfall in this argument is the fact that the Calabi-Yau
manifolds one obtains
in this way are singular. Since knot invariants get related to open
topological string
amplitudes on $Y_K$ and we already know that the disc amplitudes are not
singular, and moreover the closed string amplitudes
are expected to be the same as those for the non-singular conifold,
we do not expect the singularities to affect them.
This point, given its importance, certainly deserves further study.

The fact that $A_K(x,p;Q)$ is a mirror of resolved conifold is striking.  In particular
$A_K$ embeds as a 1-dimensional subspace of the mirror of a family of more complicated 
toric geometries.
It would be interesting to understand how the amplitudes of closed topological
strings of this more complicated geometry reduces to that of the conifold on this locus.
In this paper we have checked some aspects of this relation for some knots, which
would be nice to generalize to other knots.

The connection with knot contact homology seems interesting to explore further. The fact that knot contact homology and open Gromov-Witten theory are computing the same amplitudes was not known previously. The relation between knot contact homology and HOMFLY we uncovered is also new.
In particular it would be nice to explicitly check that the disc corrected $Q_B$ formulated
in the context of A-model open topological strings agrees with what has been considered in the
context of knot contact homology.
Moreover the results in \cite{Ng10,nget} suggest that there is another algorithm
to compute the Q-deformed A-polynomial using the braid representation of the knot.
Since the HOMFLY polynomials can also in principle be computed using the braid
representation, it would be nice to show that the large $N$ limit of HOMFLY leads
to the same result for the $A^Q_K$ as that of the knot contact homology, thus verifying this aspect of large $N$ duality
prediction.

\section*{Acknowledgments}

We would like to thank the Simons Center for Geometry and Physics where this work was initiated, during the ninth
Simons Workshop on math and physics.  We are greatly indebted to Robbert
Dijkgraaf for many discussions during various stages of this work.  We are grateful
to K. Kawagoe for computing the color HOMFLY polynomial for the figure eight knot
and sharing his unpublished work.  We are also thankful to C. Koutschan for sharing
his mathematica program for computing recursion relations.
We would also like to thank M. Culler, E. Diaconescu, T. Dimofte, P. Etingof, S. Gukov, C. Hodgson, M. Khovanov, M. Marino, L. Ng, K. Reshetikhin,  H. Segerman 
and C. Taubes for valuable
suggestions and discussions.

The research of M.A. is supported in part by the Berkeley Center for Theoretical Physics, by the National Science Foundation (award number 0855653), by the Institute for the Physics and Mathematics of the Universe, and by the US Department of Energy under Contract DE-AC02-05CH11231. The work of CV is supported in part by NSF grant PHY-0244821. 

\bibliographystyle{utcaps}	
\bibliography{myrefs}	

\end{document}